\documentclass{jfm}

\usepackage{graphicx}
\usepackage{natbib}

\usepackage{newtxtext}   
\usepackage{newtxmath}   

\newcommand{\rr}{\boldsymbol{r}}
\newcommand{\uu}{\boldsymbol{u}}

\usepackage[czech,english]{babel}
\usepackage[normalem]{ulem}

\title{Newtonian-like behavior of starting vortex flow in superfluid helium at high Reynolds numbers}

\author{J.~Blaha\aff{1}
  \corresp{Present address: Institute of Scientific Instruments, Czech Academy of Sciences, Kr\'{a}lovopolsk\'{a} 147, 612\,00 Brno, Czech Republic},
  L. Xu\aff{2}
  \and M. La Mantia\aff{1}
  \corresp{\email{lamantia@mbox.troja.mff.cuni.cz}}}

\affiliation{\aff{1}Faculty of Mathematics and Physics, Charles University, Ke Karlovu 3, 121\,16 Prague, Czech Republic
\aff{2}Department of Mathematics and Statistics, North Carolina A\&T State University, Greensboro, 27411, NC, USA}

\begin{document}
\maketitle

\begin{abstract}
We study experimentally the starting vortices shed by airfoils accelerating uniformly from rest in superfluid helium-4 (He~II).
The vortices behave apparently as if they were moving in a classical Newtonian fluid, such as air or water.
Specifically, the starting vortex positions obtained from the experimental data are found to be very close to those computed numerically in a Newtonian fluid, at sufficiently small times, when self-similar behavior is expected to occur, and for Reynolds numbers ranging approximately between $5 \times 10^2$ and $5 \times 10^5$.
The result indicates neatly that turbulent flows of He~II can be very similar to classical flows of Newtonian fluids, when thermal effects can be neglected and at sufficiently large flow scales, i.e. the study demonstrates that superfluid helium-4 could also be employed to study classical Newtonian flows.
\end{abstract}



\section{Introduction}
Superfluid helium-4, also known as He~II, is a remarkable cryogenic liquid, existing solely at temperatures lower than approximately $2$~K, in the close proximity of absolute zero \citep{mjs,bss}.
It is characterized by very small values of kinematic viscosity, which can be up to two orders of magnitude smaller than those of water \citep{donnelly98}.
This in essence means that fully developed turbulent flows, which are often associated to high Reynolds numbers, can be achieved in relatively small experimental facilities, as has been demonstrated recently in the case of turbulent vortex rings \citep{svancara22}.
However, the liquid behavior does not always resemble that of classical Newtonian fluids, such as air or water.
Indeed, as discussed e.g. by \citet{svancara19}, noticeable differences are seen when heat transport becomes non-negligible and at sufficiently small flow scales, smaller than the mean distance $\ell$ between the topological defects embedded in the fluid, which are named quantized (or quantum) vortices \citep{donnelly91}.
These vortices, which have an atomic-size core, can be of macroscopic length, i.e. they can be seen as line-like objects within the superfluid.
More importantly, they are held responsible for the unique behavior of the liquid: for example, the fact that, at sufficiently large fluid velocities, the thermal conductivity of He~II depends non-linearly on the heat flux can be related to the dynamics of quantum vortices \citep{mjs}.
The same can be said of the peculiar properties observed in some thermally driven flows of He~II, such as those in channels \citep{varga} and turbulent jets \citep{svancara23,obara}, while other flows, e.g. macroscopic vortex rings \citep{svancara20}, display distinctive Newtonian-like features also when thermally generated, i.e. the collective dynamics of quantum vortices does not result solely in non-classical characteristics when driven thermally.
Instead, mechanically generated flows of He~II display more often Newtonian-like properties at large scales, including e.g. the famous $-5/3$ energy spectrum \citep{maurer,baggaley}, the four-fifth law of turbulence \citep{salort} and intermittency, at least in some cases \citep{rusaouen,verma} -- the reader is referred to \citet{bss} for a discussion on relevant works.
Here we give our contribution to this active field of scientific research, focusing on the comparison between classical and superfluid turbulence, and present an experimental investigation that was performed in He~II, in flow conditions close to those associated to Newtonian-like behavior in previous studies, i.e. in the absence of significant heat transport and at flow scales appreciably larger than $\ell$.
Our long-term aim is to identify clear ranges of experimental parameters in which Newtonian-like features are observed in flows of superfluid helium-4.
Such ranges of experimental parameters could be instrumental in understanding why similarities between turbulent flows of He~II and analogous flows of Newtonian fluids exist in the first place, which is unknown at present.

We specifically study the starting vortices generated by airfoils accelerating uniformly from rest in He~II using the particle tracking velocimetry technique, already employed in previous studies, such as \citet{svancara21} and \citet{sakaki}.
Relatively small solid particles, suspended in the fluid, are illuminated by a thin laser sheet and their flow-induced motions are captured by a digital camera.
Then, from the positions and velocities of the particles, the vortices shed at the trailing edges of the airfoils are identified, and their relative strength is quantified using the Lagrangian pseudovorticity, a method that has already been applied to the study of large-scale vortex rings propagating in He~II \citep{svancara20,outrata}.
The starting vortex trajectories are compared to those obtained numerically in a Newtonian fluid, for a flat plate accelerating uniformly from rest \citep{xu15,xu17}, and no significant differences are observed at relatively early times.
Similarly, a satisfactory agreement is obtained from the comparison with an analytical theory, valid for an inviscid fluid \citep{pullin78,pullin04}, but only at smaller times.
That is, the agreement between experimental and numerical results is more apparent than that between experimental and analytical results.
Consequently, the studied flows of He~II display neat Newtonian-like features in the range of investigated parameters, i.e. they appear as if they were constrained mainly by viscosity, although this behavior originates most likely from the small-scale dynamics of the quantum vortices embedded in the liquid, as suggested also by recent numerical simulations on macroscopic vortex rings \citep{galantucci}.

It is now important to remark that we chose to investigate the starting flow problem in He~II because it has a rich history in Newtonian fluid mechanics, dating back to the beginning of aerodynamic theory, see \citet{reijtenbagh} and \citet{sader} for recent examples and references to some early works.
It was specifically recognized that solutions of this problem could be relevant for airfoil design and, over the years, the focus has been especially on the estimate of the forces acting on bodies accelerating from rest in a still Newtonian fluid -- these forces are in general significantly larger than those resulting from the steady motion of the same objects, as discussed e.g. by \citet{reijtenbagh}.
In other words, we chose this classical problem in view of assessing the behavior of He~II in analogous flow conditions, which were not investigated extensively in the past, although some attention has been devoted to its steady counterpart \citep{kitchens,musser}.

\section{Methods}
\begin{figure}
    \centerline{\includegraphics[width=0.6\linewidth]{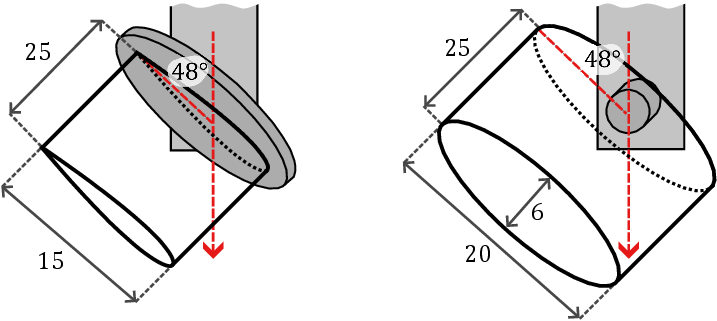}}
    \caption{Schematic views of the rectangular wings employed in the study, to scale; dimensions are in mm.
    The wing on the left side has a NACA~0012 cross-section, while that on the right side has an elliptical profile.
    The red arrows indicate the motion direction, and the angle of attack (48 degrees) is between the latter and the wing chord.
    The gray supports on the far sides of the wings link the latter to the linear motor placed outside the cryostat (they are approximately $3$~mm thick along the span direction).
    The laser sheet is located at the wing mid span, in the middle of our experimental volume -- see \citet{blaha22}, \citet{blaha24a} and \citet{blaha24b} for further technical details.
    \label{fig1}}
\end{figure}

The experiments were performed in Prague, at the cryogenic visualization laboratory of Charles University, already described in previous works, e.g. \citet{svancara21}.
Figure~\ref{fig1} shows schematic views of the two rectangular wings employed in this study.
The wing on the left side of the figure has a NACA~0012 profile, which is well-known in aerodynamics \citep[see e.g.][]{abbott}, with chord $L = 15$~mm.
That on the right side has an elliptical cross-section, with chord $L = 20$~mm and $6$~mm thick.
Both wings are made of PMMA and are $25$~mm in span.
The imposed acceleration was in both cases perpendicular to the wing span, in the vertical direction, and the angle between the motion direction and the wing chord, i.e. the angle of attack, was equal to $48$~degrees.
The camera field of view was parallel to the wing cross-section, i.e. to the wing vertical motion, and the thin laser sheet was located at the mid span, to reduce three-dimensional effects.
The wing underwent an oscillatory motion.
It first moved downward at a constant acceleration, reached a set velocity, and then decelerated to zero velocity.
After that, it moved upward in the same fashion, and came back to the initial position.
However, we are here interested in the first quarter of the cycle, when the wing is moving at constant acceleration.
For each value of acceleration $a$, approximately $100$ movies (cycles) were collected and at least $0.5$ (up to $8$) million particles' positions (and velocities) were obtained from the movies.
The data sets associated to the NACA wing are larger than those associated to the elliptical one, i.e. the former have at least 2.5 million particle positions, while the latter up to 1 million.
Further technical details are reported by \citet{blaha22} and \citet{blaha24a} for the experiments with the elliptical profile, and by \citet{blaha24b} for the NACA section experiments.

In both cases experiments were performed at constant temperature, between $1.3$ and $2.1$~K, to avoid thermal effects, and the starting vortex trajectories were found not to depend appreciably on the fluid temperature \citep{blaha22,blaha24a,blaha24b}.
Therefore, the data sets obtained at different temperatures, using the same wing and at the same acceleration magnitude, were later combined and here we do not distinguish between them, considering especially that the kinematic viscosity of He~II does not change significantly in this temperature range \citep{donnelly98}.
Following \cite{xu15}, we define the Reynolds number
\begin{equation}
    \Rey = \frac{L^2}{\nu T},
\end{equation}
where the reference time
\begin{equation}
    T= \left( \frac{L}{a} \right)^{1/2}.
\end{equation}
The kinematic viscosity $\nu$ of He~II is set here to $10^{-8}$~m$^2$~s$^{-1}$, for the sake of comparison.
The Reynolds numbers obtained in He II are then of the order of $10^5$ (see table~\ref{tab}).
To get the same $\Rey$ in water, one would need, for example, $10$~times larger chord and acceleration -- note that in the direct numerical simulation the fluid viscosity is set to $5\times10^{-6}$~m$^2$~s$^{-1}$, resulting in $\Rey \approx 500$.

\begin{table}
    \begin{center}
    \begin{tabular}{lccccc}
       & $a$ [m s$^{-2}$] & $T$ [s] & $\Rey \times 10^{-3}$ & $t_a$ & $R$  \\[3pt]
    AE & 0.17	          & 0.34	& 116.7                 & 1.46  & 1.24 \\
    BE & 0.66	          & 0.17	& 230.6	                & 1.44  & 3.20 \\
    CE & 2.59	          & 0.09	& 455.2	                & 1.42  & 8.31 \\
    AN & 0.16	          & 0.31	& 73.2	                & 1.63  & 1.01 \\
    BN & 0.63	          & 0.15	& 146.3	                & 1.63  & 2.72 \\
    CN & 2.53	          & 0.08	& 292.4	                & 1.62  & 7.30 \\
    BF & 0.64	          & 0.18	& 0.5	                & 1.41  & --   \\
    \end{tabular}
    \caption{Motion parameters.
    The motion type (first column) is indicated by two letters: the first one is associated to the acceleration magnitude $a$, with increasing value, from A to C, and the second one to the cross-section shape, with E, N and F denoting ellipse, NACA~0012 and flat plate, respectively (for the direct numerical simulation $L = 20$~mm).
    $T$ indicates the reference time and $\Rey$ is the Reynolds number;
    $t_a$ denotes the dimensionless time, in units of $T$, when the maximum profile velocity is reached, at the set acceleration;
    $R$ indicates the ratio between estimates of the smallest scale probed experimentally and the mean distance between quantum vortices (see \S\ref{results} for details).}
    \label{tab}
    \end{center}
\end{table}

As mentioned above, we estimate the starting vortex strength using the Lagrangian pseudovorticity, i.e. from the particles' positions and velocities, following \citet{svancara20} and \citet{outrata}.
This scalar quantity is defined here as
\begin{equation}
    p(\rr,t) = \left\langle \frac{\left(\rr_i - \rr\right) \times \uu_i}{|\rr_i - \rr|^2} \right\rangle,
\end{equation}
where $\rr_i$ and $\uu_i$ indicate the particle positions and velocities, respectively.
The angle brackets denote the ensemble average within a set of Lagrangian particles, captured within a time window centered at time $t$ and within an annular region centered, on a chosen grid, at the inspection point $\rr$ (the observed particle tracks lie on a plane).
The size of the annular region was chosen in order to have enough particles for the calculation of $p$ and, at the same time, to exclude contributions from particles too close to $\rr$ -- see e.g. \citet{blaha24b} for details.

It can be shown that the pseudovorticity $p$ is equal to half of the flow vorticity under ideal conditions, which are at present not met in experiments, i.e. one would need much more particles in each image to reconstruct quantitatively the vorticity field \citep{svancara20}.
Consequently, the pseudovorticity is useful to compare quantitatively different flows in the absence of Eulerian data, which is currently the case of He~II flows (the vorticity is by definition an Eulerian quantity).
Numerical simulations showed that, in the case of vortex rings, the pseudovorticity can track faithfully the ring trajectory, although its magnitude can be significantly smaller than that of the corresponding flow vorticity.
The latter was specifically observed when particles' spatial distributions typical of experiments were employed in the simulations, while particles' distributions significantly denser than those used routinely in experiments led to pseudovorticity values much closer to the vorticity ones -- see \citet{outrata} for details.

\begin{figure}
    \centerline{\includegraphics[width=1\linewidth]{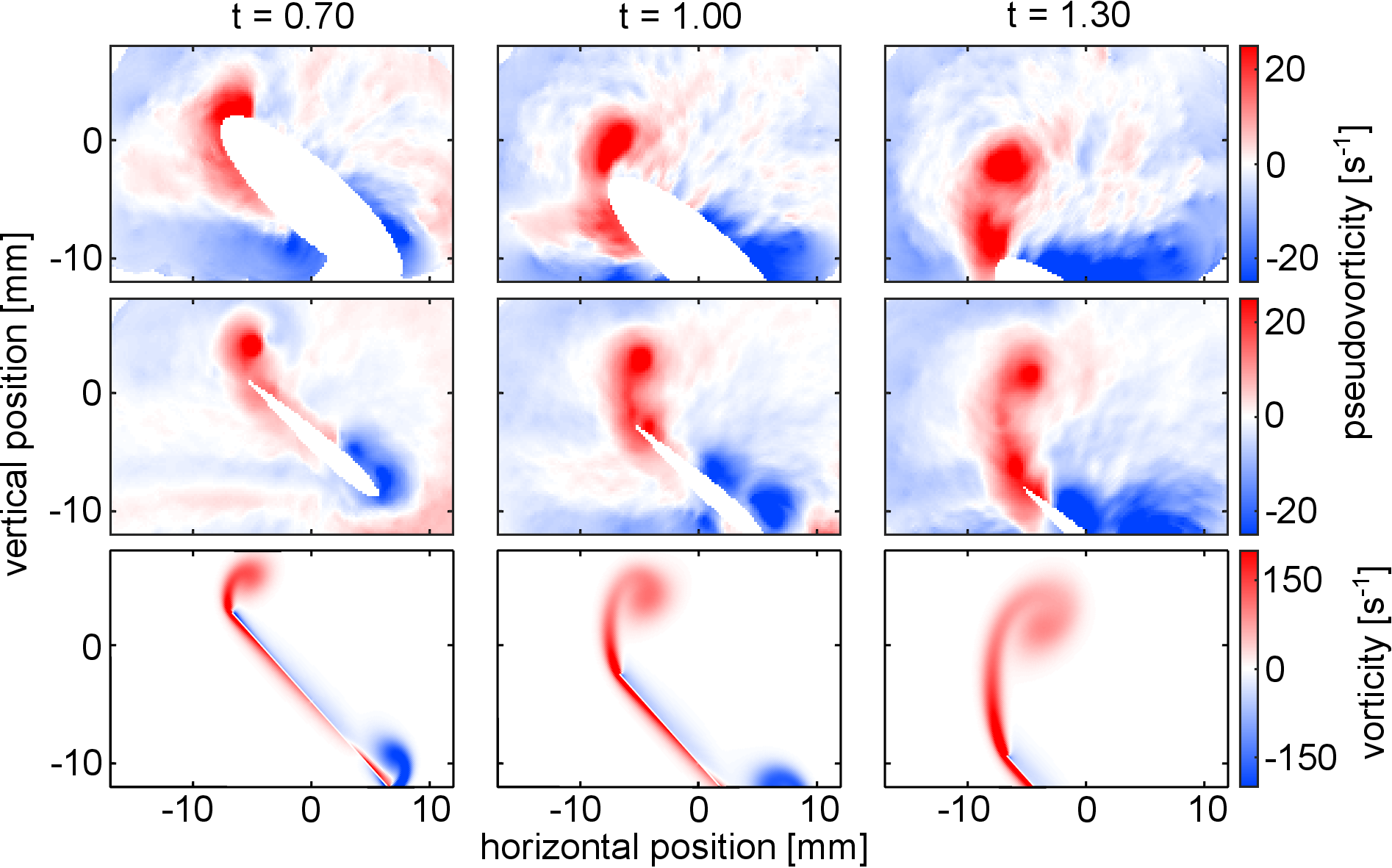}}
    \caption{Pseudovorticity and vorticity maps at three dimensionless times, in units of $T$, for the motion type BE (top row, experiment), BN (middle row, experiment), and BF (bottom row, direct numerical simulation); relevant parameters are given in table~\ref{tab}.
    The maps are plotted in the laboratory frame, with origin at the section mid chord, at $t = 0$ (at later times the wing is moving downward, in the vertical direction, at the given acceleration magnitude).
    The angle of attack (48 degrees) is between the vertical and chord directions.
    The trailing edge vortex is seen on the left side of the maps.
    \label{fig2}}
\end{figure}

\section{Results}
\label{results}
In figure~\ref{fig2} we plot the pseudovorticity and vorticity maps obtained at different dimensionless times, for three motion types.
The pseudovorticity values reported here are computed as discussed by \citet{blaha24b}, using the parameters there stated also for the ellipse motion types, for the sake of comparison.
The vorticity is computed numerically, using a fourth-order finite difference method, devised specifically to resolve the starting vortex emergence and evolution at early times, see \citet{xu15} and \citet{xu17} for technical details.
From the figure it is evident that the size of the starting vortices is strongly influenced by the profile shape, while their location seems similar.
Consequently, to investigate further the problem, in a more quantitative way, at least another step is required.

\begin{figure}
    \centerline{\includegraphics[width=1\linewidth]{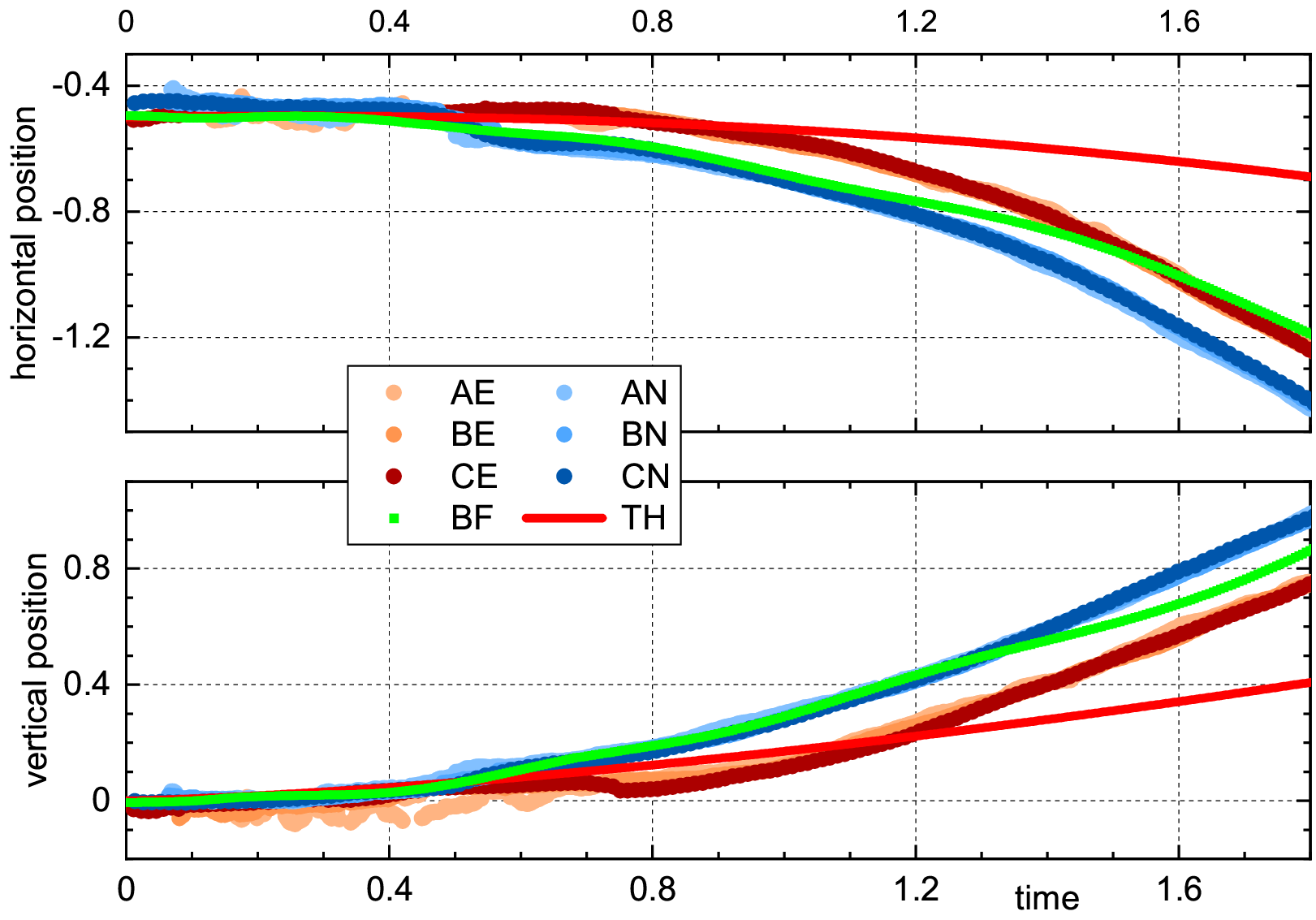}}
    \caption{Dimensionless positions, in units of $L$, of the trailing-edge vortex as a function of dimensionless time, in units of $T$, in the accelerating reference system, with origin at the profile mid chord, vertical axis perpendicular to the chord, pointing upward, and horizontal axis parallel to the chord, pointing to the right (leading edge);
    symbols as in table~\ref{tab};
    TH indicates the positions computed using an inviscid theory \citep{pullin04} considering a flat plate accelerating uniformly from rest with the chosen angle of attack.
    \label{fig3}}
\end{figure}

That is, within the obtained pseudovorticity maps we track the positions of the starting vortices shed at the profiles' trailing edges as a function of time, following the procedure described by \citet{blaha24b}, i.e. from the pseudovorticity local maxima.
The resulting positions of the starting vortices are plotted in figure~\ref{fig3}, in the accelerating reference frame; for the motion type BF (direct numerical simulation) the vortex positions were estimated using streaklines, which were also computed in previous studies \citep{xu15,xu17}.

At first sight, it is apparent that, once normalized, the vortices' trajectories depend solely on the profile shape.
Additionally, the vortices shed from sharp edges move faster than those shed from the elliptical profile.
The result could be explained considering that, in a viscous fluid, velocity gradients, which can be associated in a broad sense to energy injection into the fluid, should be more significant at a sharper edge, in comparison to a smoother one, as discussed e.g. by \citet{cadot}.

We observe a remarkable agreement between the experimental and numerical results, indicating strongly that the flows under investigation are occurring apparently in a Newtonian fluid, at least in the range of parameters here considered.
Indeed, at sufficiently small times, the vortex trajectories associated to the flat plate (BF) and to the NACA section (AN,BN,CN) are almost indistinguishable -- both profiles have a trailing edge sharper than that of the elliptical section.
The closer agreement observed at later times between the trajectories associated to the flat plate (BF) and to the elliptical profile (AE,BE,CE) can be explained considering that, at these late times, the sizes of the starting vortex and section become comparable, i.e. one could say that the symmetry of the flat plate and elliptical profile is seen by the vortex (for the NACA section the leading edge is less sharp than the trailing one).
More generally, the worsening of the agreement at later times, between the numerical and experimental results, might be due to three-dimensional, finite-size effects, which are not considered in the numerical scheme.
Note also that, as shown in table~\ref{tab}, the Reynolds number of the direct numerical simulation is much smaller than those of the experiments, i.e. our results indicate that, for the studied accelerating motions, Reynolds number effects should be negligible -- this is consistent with the numerical results reported by \citet{sader} for different types of starting flows.

We also compare the numerical and experimental results with those obtained from an inviscid theory \citep{pullin04} considering a flat plate accelerating uniformly from rest with the given angle of attack -- corresponding positions are indicated with the red line in figure~\ref{fig3}.
The agreement between the vortex positions is apparent only at relatively small dimensionless times, up to approximately $T/2$.
The result can be associated to the starting vortex self-similar growth, when the vortex size is small compared to the profile chord; at later times, when the vortex has grown into a size such that the finite section chord is no longer negligible, the inviscid model is not expected to be applicable, as discussed e.g. by \citet{xu15}.
Specifically, it has been reported that the starting vortex flow follows well the analytical self-similar theory only at some distance away from the edge (and at sufficiently small times), while viscous effects are found to be significant mainly in the close proximity of the wing, i.e. within the boundary layer, which was not accessed in the present experiments.
For instance, \citet{pullin80} conducted relevant experiments in water and compared satisfactorily the obtained results with those of an inviscid fluid theory \citep{pullin78} -- see \citet{sader} for other relevant examples.

\begin{figure}
    \centerline{\includegraphics[width=1\linewidth]{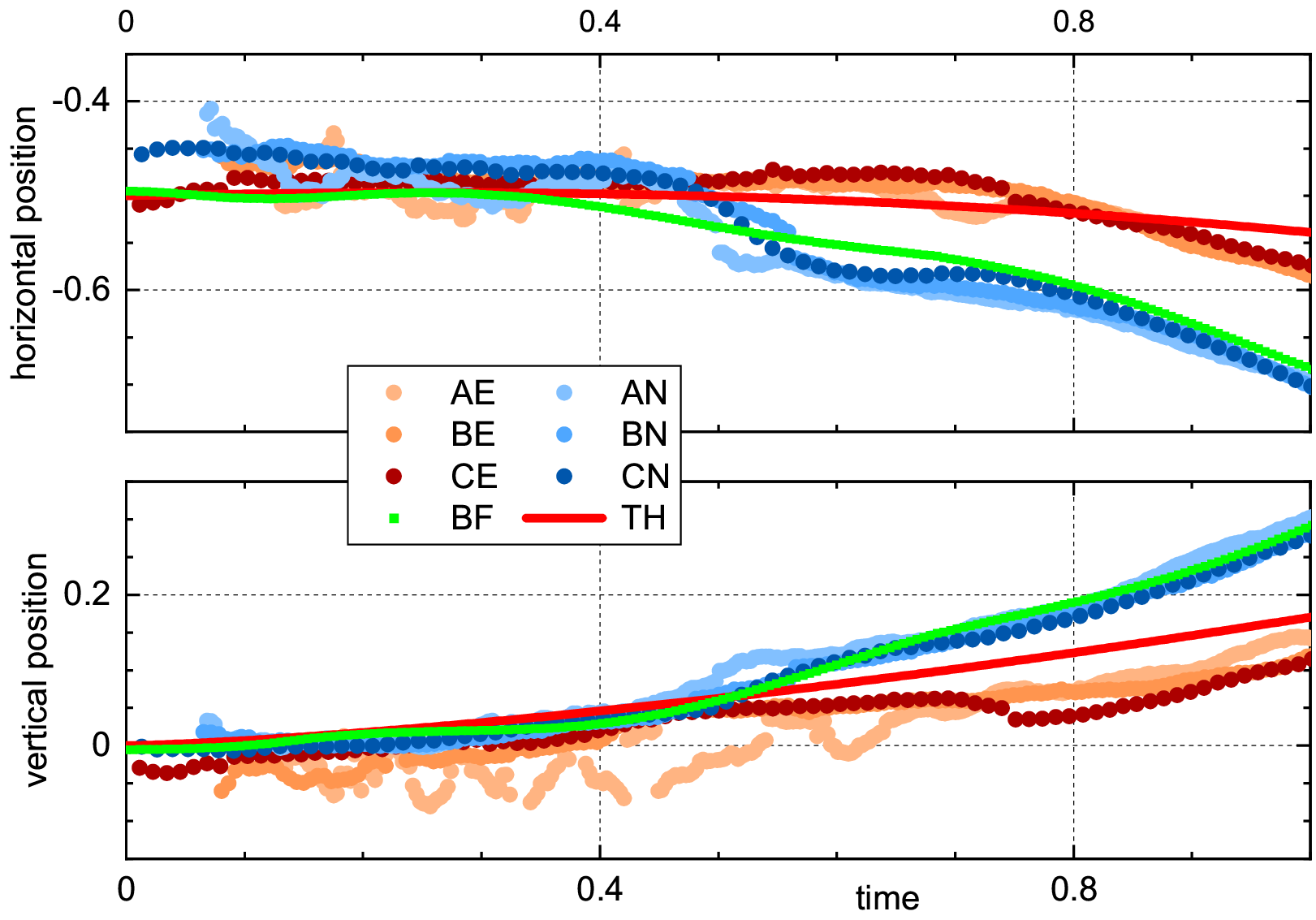}}
    \caption{Enlargement of figure \ref{fig3} at early times;
    symbols as in figure~\ref{fig3}.
    \label{fig4}}
\end{figure}

Indeed, as shown in figure~\ref{fig4}, the agreement between the numerical and inviscid theory results is more evident at the smallest times, when self-similar behavior is expected to occur.
It is also apparent from the figure that, at these small times, the vortex positions identified from the experimental data are more scattered than those computed from the inviscid theory and the numerical data, although the overall trends are consistent with each other, especially at the highest accelerations.
To improve the situation, one would likely need much more particles in each image, to identify more neatly the pseudovorticity local maxima in the proximity of the wing.
Note, in this regard, that the positions associated to the elliptical profile are more scattered than those related to the NACA section, which were obtained using significantly larger data sets, as noted above.

\begin{figure}
    \centerline{\includegraphics[width=0.7\linewidth]{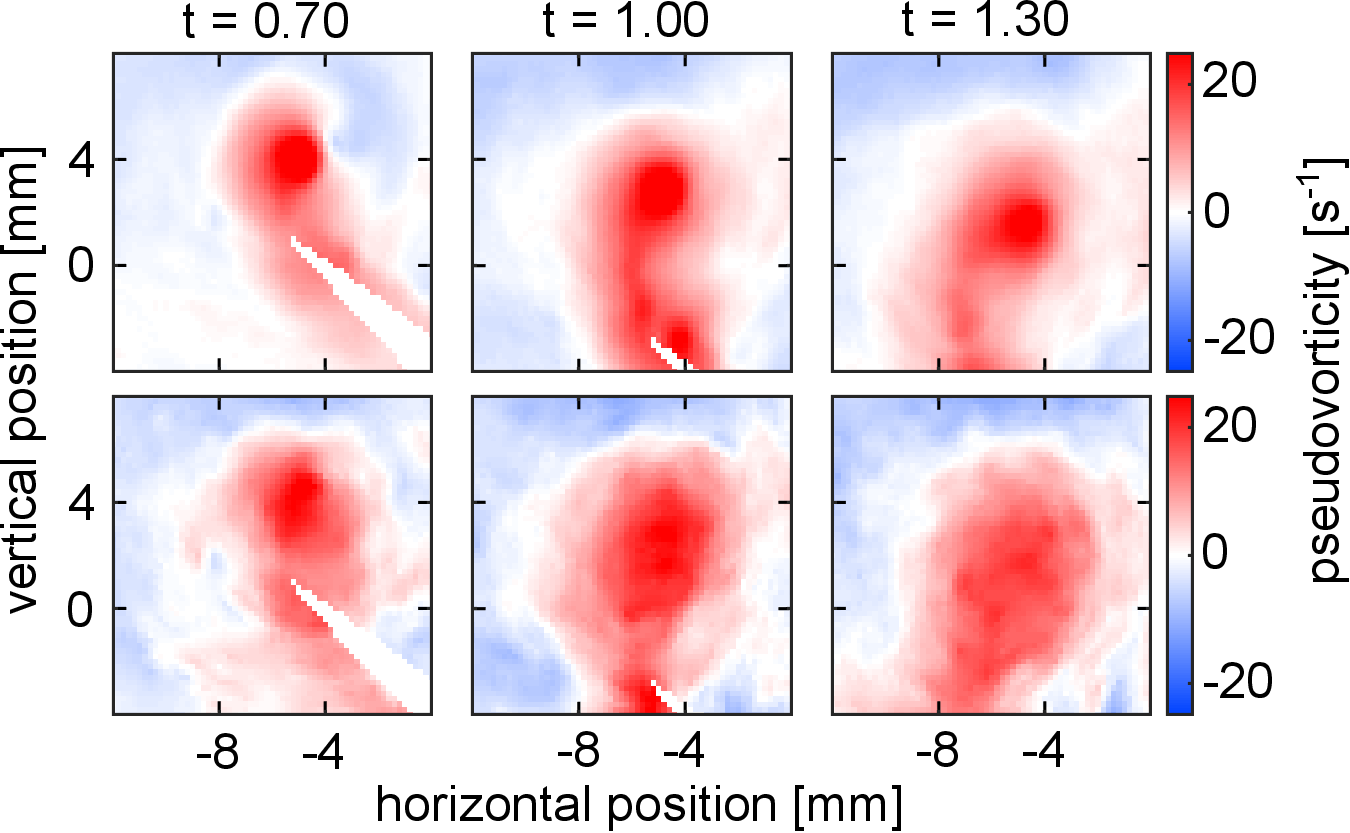}}
    \caption{Top row: pseudovorticity maps at three dimensionless times, in units of $T$, for the motion type BN, see table~\ref{tab} for relevant parameters.
    Bottom row: pseudovorticity maps at three dimensionless times, as in the top panels, for a motion type similar to BN, but occurring in a Newtonian fluid (He~I), with $a=0.63$~m~s$^{-1}$, $T=0.15$~s, $\Rey = 73 \times 10^3$ -- the corresponding kinematic viscosity $\nu$ is equal to $2\times10^{-8}$~m$^2$~s$^{-1}$ \citep{donnelly98} -- and $t_a = 1.62$.
    The maps are plotted in the laboratory frame, as in figure~\ref{fig2}, but the axes extrema are smaller than those of figure~\ref{fig2}, to highlight the starting vortex location.
    \label{fig5}}
\end{figure}

It is now useful to say that we also performed similar experiments in a Newtonian fluid, using the same NACA wing and the same visualization technique.
Specifically, we chose parameters close to those associated to the motion type BN, see table~\ref{tab}.
The main difference is the Reynolds number, i.e. these experiments were performed in liquid helium-4, at $2.3$~K, which is a classical Newtonian fluid, called helium I or He I.
This liquid is characterized by relatively small values of kinematic viscosity, up to three times higher than those of He~II \citep{donnelly98}, and by values of thermal conductivity that can be orders of magnitude lower than those of He~II \citep{mjs} -- note in passing that He~I becomes He~II (superfluid helium-4) at approximately $2.2$~K \citep{donnelly98}.
The number of particles' positions (and velocities) obtained in He~I was approximately five times smaller than in He~II, because fewer movies were collected for these parameters.
The resulting pseudovorticity maps are consequently more scattered in He~I than in He~II, as is apparent in figure~\ref{fig5}.
It also follows that the vortex tracking algorithm employed here \citep{blaha24b} is not suitable for such scatter data sets.
However, from these maps one can still see that the location of the starting vortex looks similar in these two fluids, as in figure~\ref{fig2}.
Additionally, we performed more recently another series of experiments, using the same wing and visualization technique, both in He~I and in He~II, with analogous motion parameters, and no appreciable differences are observed in the starting vortex trajectories.
In this more recent experimental campaign we collected more data in He~I, in comparison to the present experiments, but the angle of attack was slightly larger (49 degrees) and three-dimensional effects were more evident -- see \citet{brichet} for preliminary results.

In summary, our main outcome, shown in figure~\ref{fig3}, indicates clearly that the flows of superfluid helium-4 here investigated behave as if they were flows of a Newtonian fluid.
To substantiate further the claim, it is useful to mention that numerical results obtained employing the same computational techniques have already shown agreement with relevant experimental observations, as discussed e.g. by \citet{xu15}.
Similarly, it was already shown that, under some conditions, experimental results obtained in Newtonian fluids cannot be distinguished from those obtained in He~II employing the same investigative tools, as demonstrated e.g. by \citet{maurer} and \citet{svancara17}.
In the same spirit of these studies, we now estimate the mean distance $\ell$ between quantum vortices to show that here we are actually probing flow scales significantly larger than $\ell$ because, as noted above, we do expect Newtonian-like features to occur at these large flow scales.
Specifically, in the absence of significant thermal effects, as in the present case, one can write, following e.g. \citet{blaha24b},
\begin{equation}
    \ell = \left( \frac{\kappa}{\omega} \right)^{1/2},
\end{equation}
where $\kappa=9.97 \times 10^{-8}$~m$^2$~s$^{-1}$ is the quantum of circulation \citep{donnelly98}, associated to each quantized vortex, and $\omega$ denotes the flow vorticity.
Additionally, as noted above, it was shown that the pseudovorticity is equal to half of the vorticity under conditions that are at present not met in experiments \citep{svancara20}.
Consequently, to obtain an upper-bound estimate of the mean distance between quantum vortices, we set
\begin{equation}
    \ell = \left( \frac{\kappa}{2 \overline{p}} \right)^{1/2},
\end{equation}
where $\overline{p}$ indicates the average value of the positive pseudovorticity, associated to the trailing edge vortex and computed from the experimental data, for dimensionless times smaller than $1.8$, which is consistent with the time range displayed in figure~\ref{fig3}.
Similarly, an estimate of the smallest scale probed by our particles can be obtained multiplying the maximum profile velocity, reached at the dimensionless time $t_a$ reported in table~\ref{tab}, and the time between subsequent particle positions, used for the particles' velocity estimate ($0.002$~s).
The ratio $R$ between the latter and the estimate of the mean distance between quantum vortices is listed in table~\ref{tab}.
Considering that, as already noted, in the case of vortex rings \citep{outrata} the pseudovorticity can be significantly smaller than the vorticity, we can safely say that in the present case our particles are probing flow scales significantly larger than the mean distance between quantum vortices.
Note, in this regard, that the magnitudes of the vorticity maxima obtained in the direct numerical simulation (BF) are significantly larger than those in the pseudovorticity maps, as shown in figure~\ref{fig2}.

\section{Conclusions}
The results reported in the present work demonstrate striking similarities between starting vortex flows in He~II and in a Newtonian fluid, at sufficiently large flow scales, larger than the mean distance between quantum vortices, and in the absence of significant thermal effects.
Is this just a coincidence, or could quantitative relations exist between the collective behavior of quantized vortices and viscosity effects?
We believe that further studies are required to clarify the issue, e.g. future investigations on starting vortex flows should be devoted to collect more data not only in He~II but also in Newtonian fluids, especially in view of comparing the shape and strength of the shed vortices.
More generally, we believe that the quantitative identification of ranges of experimental parameters in which Newtonian-like features are observed in flows of He~II could be instrumental in understanding why these similarities exist in the first place.
On the other hand, these similarities could enable us to investigate classical flows directly using He~II, within the identified parameters' ranges, taking especially advantage of the extremely small kinematic viscosity of this unique liquid, resulting in relatively small experimental facilities, significantly smaller than typical wind tunnels.
Indeed, the latter have cross-section areas of $O(10^4$)~cm$^2$, while the cross-section of the cryogenic wind tunnel employed in the present experiments is 25~cm$^2$~\citep{blaha24b}.

\backsection[Acknowledgements]{L.~Xu and M.~La~Mantia thank M.~Nitsche and R.~Krasny for fruitful discussions.
M.~La~Mantia thanks P.~Dabnichki, for providing the PMMA wings employed in the study, M.~Rotter, for valuable help during the experiments' preparation, and L.~Brichet, for contributing to the more recent experimental campaign mentioned in \S\ref{results}.}

\backsection[Funding]{J.~Blaha and M.~La~Mantia acknowledge financial support from the Ministry of Education, Youth and Sports of the Czech Republic under Grant No.~LL2326.}

\backsection[Declaration of interests]{The authors report no conflict of interest.}

\backsection[Data availability statement]{The data that support the findings of this study are available from the corresponding author upon reasonable request.}

\backsection[Author ORCIDs]{J.~Blaha, https://orcid.org/0009-0001-0774-645X; L.~Xu, https://orcid.org/0000-0002-9971-9888; M.~La~Mantia, https://orcid.org/0000-0002-7159-5924.}

\backsection[Author contributions]{M.~La~Mantia designed research.
Experiments were performed in Prague by J.~Blaha and M.~La~Mantia.
Numerical simulations were performed by L.~Xu.
All authors contributed to data processing and interpretation, and to manuscript preparation.}

\bibliographystyle{jfm}
\bibliography{biblo}

\end{document}